# The mass composition of cosmic rays above $10^{17}$ eV*

**A A Watson**
**School of Physics and Astronomy**
**University of Leeds**
**Leeds LS2 9JT, UK**

**Abstract:** It is shown that our knowledge of the mass composition of cosmic rays is deficient at all energies above $10^{17}$ eV. Systematic differences between different measurements are discussed and, in particular, it is argued that there is no compelling evidence to support the common assumption that cosmic rays of the highest energies are protons. Our knowledge of the mass needs to be improved if we are to resolve uncertainties about the energy spectrum and interpret data on the arrival direction distribution of cosmic ray.

**1. The Scientific Motivation for Studying the Highest Energy Cosmic Rays:** Since the recognition in 1966, by Greisen and by Zatsepin and Kuzmin, that protons with energies above $4 \times 10^{19}$ eV would interact with the cosmic microwave radiation, there has been great interest in measuring the spectrum, arrival direction distribution and mass composition of ultra high-energy cosmic rays (UHECR), defined as those cosmic rays having energies above $10^{19}$ eV. Specifically, it was pointed out that if the sources of the highest energy protons were universally distributed, then there should be a steepening of the energy spectrum in the range from 4 to $10 \times 10^{19}$ eV. This predicted feature has become known as the GZK cut-off but the sharpness of the steepening depends on unknown factors such as the evolution and production spectrum of the sources. If the UHECR are mainly Fe nuclei then there will also be a steepening, but it is harder to predict the character of this feature as the relevant diffuse infrared photon field is poorly known: steepening is expected to set in at higher energy.

Early instruments built to explore this energy region (at Volcano Ranch (USA), Haverah Park (UK), Narribri (Australia) and Yakutsk (USSR)), were designed long before the 1966 predictions and when the flux above $10^{19}$ eV was poorly known. Although of relatively small area (~10 km²) sufficient exposure was eventually accumulated to measure the rate of cosmic rays above $10^{19}$ eV accurately and to give the first indications that there might be cosmic rays with energies above $10^{20}$ eV, well above the GZK cut-off. Evidence for anisotropies above $10^{19}$ eV was not established. Over the same period, it also came to be accepted that the problem of acceleration of protons and nuclei to such energies in known astrophysical sources is a major one[1].

The exposures that were achieved with these observatories was too small to answer the question of the spectral steepening in a convincing manner, though had the CMB temperature been 3.3 K rather than 2.7 K, the outcome might have been different. The projects that followed (Fly's Eye, AGASA and HiRes) also gave indications of trans-GZK particles but by the early 1990s it was apparent that even areas of 100 km² operated for many years could not measure the properties of UHECR in adequate detail. In 1991, Cronin and Watson began the design of a suitable instrument and the task of assembling an international collaboration to fund and construct it. The collaboration and funding appeared sufficiently robust for work on the Pierre Auger Observatory to begin in Malargüe, Mendoza Province, Argentina, in March 1999. This southern observatory is seen as the first of the two that are needed to provide full sky coverage.

**2. Importance of mass composition for accurate estimates of the energy spectrum:** The energy spectrum of cosmic rays has frequently been inferred from observations of signals in scintillators or water-Cherenkov detectors deployed at altitudes not far from sea-level (see Nagano and Watson[2] for a relevant review). In practice, the energy is derived from a measurement of the detector signal at (typically) 600 m from the shower axis using the results of detailed Monte Carlo calculations. For example, in the most recent reports of spectra from the AGASA[3] and Haverah Park groups[4], the QGSJET model has been adopted. For the AGASA work, a study of the impact of different models has been made and it is found using QGSJET98, for example, in the Corsika propagation code, that the energy estimates for protons or iron nuclei differ by about 12% near $10^{20}$ eV (with protons giving the higher energy). With the SIBYLL 1.6 model the difference is 19%, in the same direction and the energy estimates are about 5% higher. Thus, these two sources of systematic error in the energy estimates – lack of knowledge of the mass and uncertainty about the model – are essential to keep in mind when comparing different data sets. As the relevant centre-of-mass energies at $10^{20}$ eV are about 30 times that of the LHC, it is clear that the systematic effect in the model is extremely difficult to quantify.

The other method that has been used to measure primary energies is the fluorescence technique[5] in which photons emitted by excited $N_2$ are observed. With this method, it is possible to make a calorimetric estimate of the large fraction of the primary energy that is transferred into ionisation as the shower particles cross the atmosphere. The estimate of this part of the primary energy loss is relatively model independent with, as first discussed by Linsley[6], and more recently by Song et al.[7], a correction of around 10% to be made for energy that does not go into atmospheric ionisation.

With these remarks in mind, it is useful to discuss the spectra reported by the HiRes[5] and AGASA[3] groups. A convenient plot of the data from these experiments has been prepared by Bergman[8] and is shown in figure 1. A log-log plot of J (the differential intensity) against energy, E, has significant merits over the $JE^3$ vs. E plots commonly adopted in recent years. In particular, propagators of the latter form (including the present author) almost always ignore the fact that error bars should be shown as diagonal lines and that the uncertainty in energy (often justifiably omitted in the x-direction because of the bin size) should be added in quadrature to the uncertainty in intensity, as it comes in as the third power. $JE^3$ vs. E representations do the data a disservice as the incorrect error assignments serve to give an erroneous impression of the level of incompatibility between the data sets.

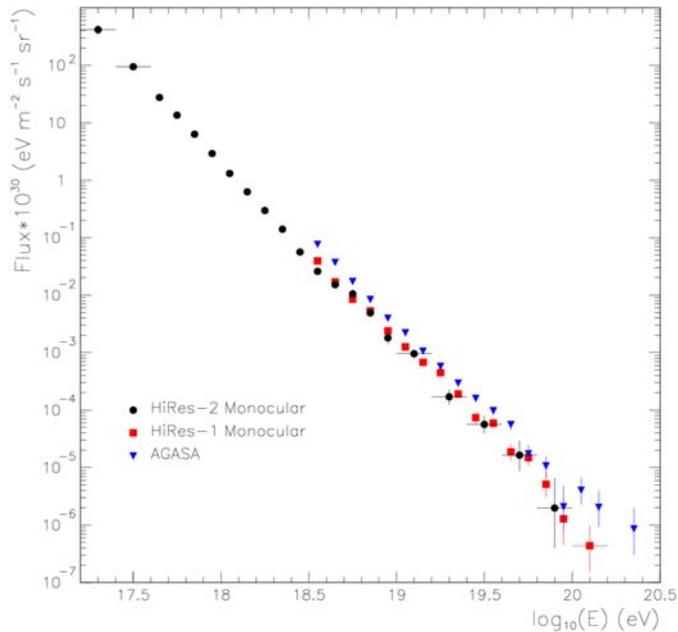

**Figure 1:** The energy spectra as reported by the AGASA[3] and HiRes[5] groups. This clear presentation of the spectra is due to D Bergman (Rutgers University).

It is clear that the spectra reported by AGASA and HiRes, and shown in figure 1, could be largely reconciled if the energy scale of one or other was adjusted by 30%, or if each was moved by 15%. Moreover, the possibility that there are uncertainties in the flux measurements should not be overlooked. At the lower end of the AGASA spectrum the aperture is changing quite rapidly with energy[3] and uncertainties in the lateral distribution function, that describes the fall-off of signal with distance, may lead to uncertainties in the aperture determination. To calculate the sensitive area requires knowledge of the lateral distribution function that is difficult to obtain experimentally. Estimates from Monte Carlo calculations are not reliable because of model and mass uncertainties. Thus, the differences between HiRes and AGASA near 3 EeV may arise because of uncertainties in the flux and in the mass used to estimate the energy. At the highest energies, only the issues of model and mass are important for the AGASA Observatory as events included in the spectrum all have cores that lie within the boundary of the array.

By contrast, with fluorescence detectors, the aperture continues to grow with energy and there remains considerable uncertainly about the HiRes aperture. Extensive Monte Carlo calculations must be made to establish it and these make assumptions about the slope of the spectrum and about the primary mass, with protons being assumed on the basis of claims from the Fly's Eye and HiRes experiments[9], ignoring the often contradictory evidence from other experiments without discussion. Further, in the data published so far by the HiRes group,s corrections for atmospheric effects on a night by night basis have not yet been included: rather an average atmosphere is used. It will be possible to interpret the aperture with less ambiguity for the HiRes stereo data as the core position can be determined by geometry. Note that with the Auger Observatory, all of the fluorescent events will have surface detector signals so that the cores (and aperture) are well-constrained.

Despite these caveats, it appears certain that trans-GZK events do exist. The Utah group reported an event of 300 EeV in 1993[10] from the Fly's Eye detector, the AGASA group have claimed an event of 210 EeV[11] together with several other events with energies reported above 100 EeV, while a stereo HiRes event has been reported at 220 EeV[12]. We understand that this latter event is not included in the HiRes spectra of figure 1 as it was recorded during a short period of good atmospheric conditions on a night that was otherwise rather unstable.

What is presently in doubt is *not* the *existence* of trans-GZK events but rather the *flux* of them.

**3. The mass of UHECR:** Our knowledge about the mass of primary cosmic rays at energies above $10^{17}$ eV is rudimentary. Different methods of measuring the mass give different answers and the conclusions are dependent, to a greater or lesser extent, upon the model calculations that are assumed. Results from some of the techniques that have been used are now described and the conclusions drawn reviewed. Some of these techniques will be applicable with the Pierre Auger Observatory[13]. Systematic errors that arise in measurements and models remain a problem.

**3.1. The Elongation Rate:** The elongation rate is the term used to describe the rate of change of depth of shower maximum with primary energy. The term was introduced by Linsley[14] and, although his original approach has been, to some extent, superseded by the results of detailed Monte Carlo studies, the concept remains useful for organising and describing data and calculations. Figure 2 shows a summary of measurements of the depth of maximum together with predictions from a variety of model calculations[15]. It is clear, if certain models are correct, that one might infer that the primaries above $10^{19}$ eV are dominantly protons but that others are indicative of a mixed composition. In particular, the QGSJET set of models (basic QGSJET01 and the 5 options discussed in [15]) and the Sibyll 2.1 model force contrary conclusions. Note that the experimental data on $X_{max}$ do not yet extend beyond 20 EeV.

**3.2. Fluctuations in Depth of Maximum:** Further insight has long been expected to come from the magnitude of fluctuation of the position of depth of maximum. If a group of showers is selected with a narrow range of energies, then fluctuations about the mean of $X_{max}$ are expected to be larger for protons than for iron nuclei. A recent study of this has been reported by the HiRes group[9] for 553 events above 10 EeV. It is argued, using the Sibyll or QGSJET models, that the fluctuations are so large that a large fraction of protons must be present in the primary beam, in apparent agreement with inferences sometimes drawn from figure 2. However, the HiRes data have been analysed assuming a standard atmosphere for each event. This is unlikely to represent reality, as it is probable that the atmosphere deviates from the standard conditions from night to night and even during a night of observation. This view is strengthened by the results of balloon flights made from Malargüe[16], which have shown that the atmosphere changes in a significant way both diurnally and seasonally. If a standard atmosphere is used, some of the fluctuations observed in $X_{max}$ may be incorrectly attributed to shower, rather than to atmospheric, variations.

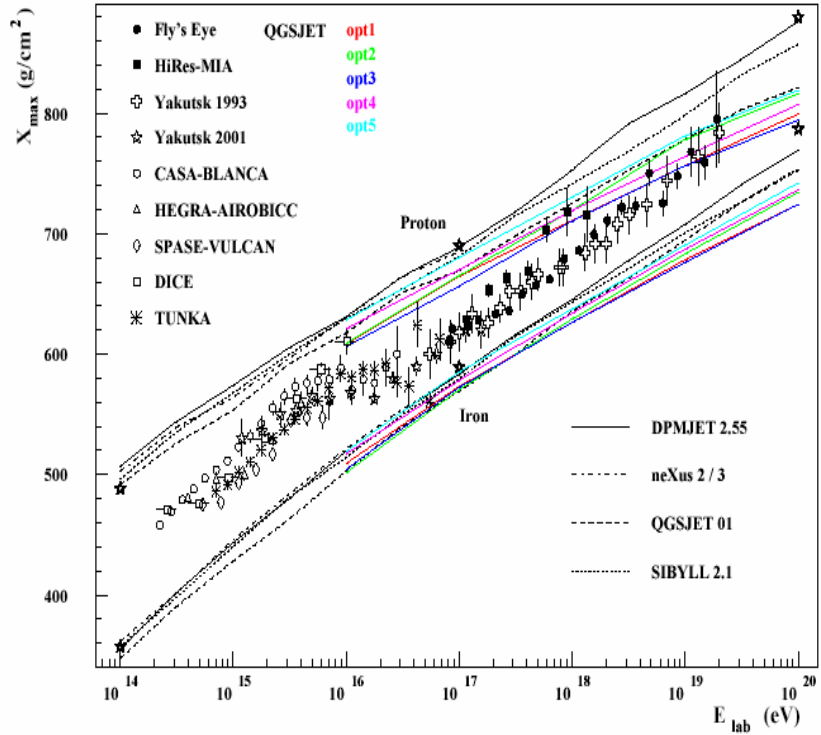

**Figure 2:** The depth of maximum, as predicted using various models, compared with measurements. The predictions of the five modifications of QGSJET, discussed in [15], from which this diagram is taken, lie below the dashed line that indicates the predictions of QGSJET01.

Doubts about traditional interpretations are reinforced by the detailed Monte Carlo analysis of uncertainties in $X_{max}$ and the fluctuations in $X_{max}$ discussed in these proceedings by L Perrone[17]. It is found that at distances beyond 20 km, there are significant systematic shifts in the $X_{max}$ values derived and in the spread of the $X_{max}$ values. At 20 km, $X_{max}$ is, on average, estimated to be 60 g cm$^{-2}$ deeper in the atmosphere and the fluctuations in $X_{max}$ for iron nuclei are considerable with $\sigma \sim 100$ g cm$^{-2}$. These factors act in such a way as to suggest that the elongation rate reported by HiRes (and presumably also by Fly's Eye) may have been systematically over-estimated and that the fluctuations in $X_{max}$ are not due entirely to protons. Thus, it may be premature to draw conclusions about the presence of protons from analyses of fluctuations.

**3.3. Mass from muon density measurements:** It is well known that a shower produced by an iron nucleus will contain a larger fraction of muons at the observation level than a shower of the same energy created by a proton primary. Many efforts to derive the mass spectrum of cosmic rays have been made using this fact. However, although the differences are predicted to be relatively large (on average there are ~70% more muons in an iron event than a proton event), there are large fluctuations and, again, there are differences between what is predicted by particular models. Thus, the QGSJET set predicts more muons than the Sibyll family, the difference arising from different predictions as to the pion multiplicities produced in nucleon-nucleus and pion-nucleus collisions that in turn arise from differences in the assumptions about the parton distribution within the nucleon[18]. In a contribution to these proceedings, K Shinosaki[19] has described the data on muons signals from the AGASA array.

There are 129 events above $10^{19}$ eV, of which 19 have energies greater than 3 x $10^{19}$ eV. Measurements of muon densities at distances between 800 and 1600 m were used to derive the muon density at 1000 m with an average accuracy of 40%. This quantity is compared with the predictions of model calculations. The difference between the proton and iron predictions is small, especially when fluctuations are considered. The AGASA group concludes that above $10^{19}$ eV the fraction of Fe nuclei is < 40% at the 90% confidence level. In my view, the 5 events above $10^{20}$ eV for which such measurements are possible, are fitted as well by iron nuclei as by protons. Further, the conclusions are sensitive to the model used: as the Sibyll model predicts fewer muons than the QGSJET model, higher iron fractions would have been inferred had that model been adopted.

At lower energies, there are muon data from the Akeno array and from AGASA[20]. Different analyses have been made of these. The AGASA group claims that the measurements are consistent with a mass composition that is unchanging between $10^{18}$ and $10^{19}$ eV.

**3.4. Mass estimates from the lateral distribution function:** The rate of fall of particle density with distance from the shower axis provides another parameter that can be used to deduce the mass composition. Showers with steeper lateral distribution functions (LDFs) than average will arise from showers that develop later in the atmosphere, and vice versa. A detailed measurement of the LDFs of showers produced by primaries of energy greater than $10^{17}$ eV was made at Haverah Park using a specially constructed 'infilled array' in which 30 additional water tanks of 1 m$^2$ were added at the centre of the array on a grid with spacing of 150 m. When the work was completed in 1978, the data could not be fitted with the shower models then available, for any reasonable assumption about the primary mass. Recently[21], this data set has been re-examined using the QGSJET98 model. The choice of this model was justified by showing that it adequately described data on the time spread of the Haverah Park detector signal over a range of zenith angles and distances near the core (<500 m). Here the difference predicted between the average proton and iron shower is only a few nanoseconds and the fit achieved is good. Density data were fitted by a function $\rho(r) \sim r^{-(\eta + r/4000)}$, where $\eta$ is the steepness parameter. The spread of $\eta$ was compared with predictions for different primary masses. The proton fraction, assuming a proton-iron mixture, is found to be independent of energy in the range 3 x $10^{17}$ to $10^{18}$ eV and is (34±2) %. If this fraction is evaluated with QGSJET01, in which a different treatment of diffractive processes is adopted from that in QGSJET98, then the fraction increases to 48%. The fraction is larger because the later model predicts shower maxima that are higher in the atmosphere and accordingly, to match the observed fluctuations, the proton fraction must be increased. The difference in the deduced ratio thus has a systematic uncertainty from the models that is larger than the statistical uncertainty. Although the necessary analysis has not been made, it is clear that the Sibyll 2.1 model would be consistent with a smaller fraction of protons.

A similar analysis has been carried out using data from the Volcano Ranch array[22]. As with the Haverah Park information, no satisfactory interpretative analysis was possible when the measurements were made. Using 366 events for which Linsley left detailed information, and QGSJET98, the fraction of protons is estimated as (11 ± 5(stat) ± 12 (sys))% at ~ 1 EeV [17]. With QGSJET01 the flux of protons would increase to ~ 25%, indicating again that model uncertainties remain a serious barrier to interpretation.

**3.5. Mass from the thickness of the shower disk:** The particles in the shower disc do not arrive at a detector simultaneously, even on the shower axis. The arrival times are spread out because of geometrical effects, velocity differences, and because of delays caused by multiple scattering and geomagnetic deflections. The first particles to arrive (except very close to the shower axis) are the muons as they are scattered rather little and geometrical effects dominate. At Haverah Park four detectors, each of 34 m$^2$, provided a useful tool for studying the thickness of the shower disc, which will depend upon the development of the cascade. Recently, an analysis of 100 events of mean energy ~$10^{19}$ eV has shown that the magnitude of the risetime is indicative of a large fraction (~80%) iron nuclei at this energy[23]. It is expected that this type of study will be considerably extended with the Pierre Auger Observatory, in which the photomultipliers each water tank are equipped with 25 ns flash ADCs.

### 3.6 Summary of data on primary mass above 0.3 EeV

In figure 3, taken from [22], the results taken from various reports of the Fe fraction are shown. It is disappointing that the data from Volcano Ranch and from Haverah Park are not in better agreement as a similar quantity, the lateral distribution function of the showers, was measured at each array and the same model - QGSJET98 - was used to interpret the data, although with different propagation codes (AIRES and CORSIKA respectively). We cannot explain this difference: at $10^{18}$ eV the estimates of the fraction of Fe are separated by over 2 standard deviations.

In figure 3, there are also data from the Akeno/AGASA and the Fly's Eye experiments. The Akeno/AGASA groups measured the muon densities in showers, normalised at 600 m. The energy thresholds for Akeno and AGASA were 1 and 0.5 GeV respectively. The Fly's Eye data are deduced from measurements of the depth of shower maximum. In an effort to reconcile differing claims made by the two groups of the trend of mass composition with energy, Dawson et al.[24] reassessed the situation used a single model, SIBYLL 1.5 to interpret both data sets. SIBYLL 1.5 was an early version of the SIBYLL family that evolved to SIBYLL 1.6 and 1.7, and most recently, to SIBYLL 2.1[18]. It is the estimates of the Fe fractions from [24] that are shown in figure 3. There are major discrepancies between these estimates and between those from Volcano Ranch and Haverah Park. However, the predictions of the muon density and of the depth of shower maximum made with the version of SIBYLL used differ significantly from those that would be derived now using QGSJET98 or 01 (or with SIBYLL 2.1). We now discuss this point in some detail.

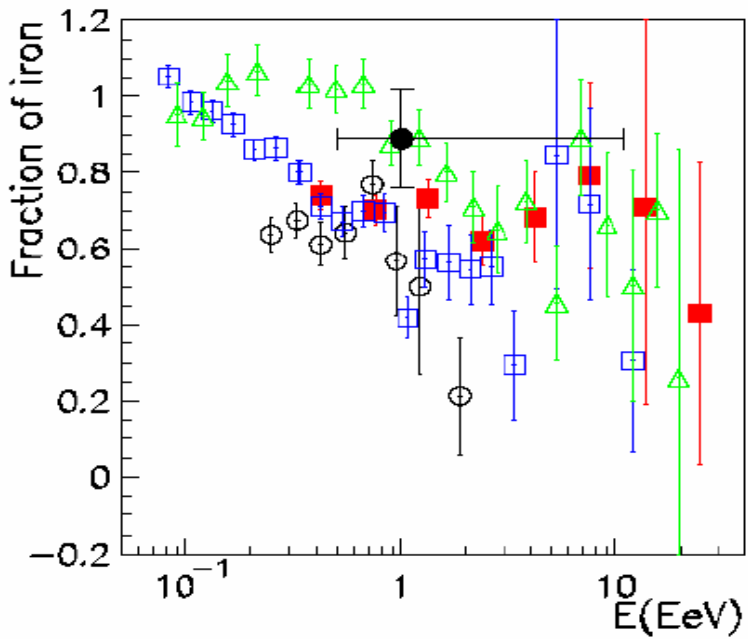

**Figure 3:** The fraction of Fe nuclei as a function of energy as reported from various experiments. Fly's Eye ($\triangle$), AGASA A100 ($\blacksquare$), AGASA A1 ($\square$), Haverah Park ($\bigcirc$) and Volcano Ranch ($\bullet$) (Figure taken from [22]).

An extremely useful set of comparisons of the predictions from SIBYLL 1.7 and 2.1 with those from QGSJET98 has been given in [18]. We understand that SIBYLL 1.6 and SIBYLL 1.7 differ only in that the neutral pions were allowed to interact in the latter model and it is not believed that this will make a serious difference to the predictions at energies below $10^{19}$ eV[25]. Therefore, in what follows, we regard the SIBYLL 1.7 and the QGSJET98 differences as being identical to those that exist between SIBYLL 1.6 (or 1.5) and QGSJET98, for which no similar comparisons are available. It is convenient to make comparisons at $10^{18}$ eV. More detailed cross-checks, over a range of energies, would require more extensive knowledge of features of the Fly's Eye and Akeno/AGASA systems than we possess.

Turning first to the data from the depth of maximum, we note that at $10^{18}$ eV the measured value of $X_{max}$ is ~675 g cm$^{-2}$, with an error that is less than the size of the data point (< 10 g cm$^{-2}$). The predictions for proton primaries made with SIBYLL 1.7 and QGSJET98 are 760 and 730 g cm$^{-2}$ respectively[18]. Thus, a mass composition less dominated by Fe is favoured compared with the ~90% estimated in [24]. The choice of SIBYLL 2.1 would alter this argument rather little as the predicted depth[18] at $10^{18}$ eV is 740 g cm$^{-2}$. Further study of this matter could be made but the data from Fly's Eye should soon be superseded by more definitive data from the HiRes stereo system and from the Auger instruments.

A qualitative statement about the shift expected in the Fe fraction, estimated from the measurement of muon densities at 600 m, that will come from changes in the model can be made using information in [18]. Although the calculations do not exactly match the energies of the Akeno/AGASA measurements (> 0.3 GeV is computed and > 0.5 GeV measured), ratios between the predictions of different models are not strongly dependent upon energy threshold. What is of importance is the ratio of the number of muons predicted, at $10^{18}$ eV, for SIBYLL 1.7, SIBYLL 2.1 and QGSJET98. At $10^{18}$ eV, these numbers are in the ratios 1: 1.17: 1.44. The difference in muon number between SIBYLL 1.7 and QGSJET98 is comparable to that expected between proton and Fe primaries (~50%, but also model dependent). It is clear that the more recent models, if applied to the Akeno/AGASA data after the manner of the analysis of [24], would lead to a significant reduction in the predicted fraction of Fe nuclei. To pursue this further would require knowledge of the predicted densities at 600 m, information that is presently lacking. We note that the shift in the Fe fraction from the muon data is probably substantially larger than it is when using the data on $X_{max}$.

We are not able to use the information reported from the HiRes-MIA experiment[26] in which muons and $X_{max}$ were observed simultaneously. As with Akeno/AGASA, the muon density at 600 m was determined. The problem we have is that while the papers describe the data as being consistent with a mass composition that becomes lighter with energy, this appears, on close scrutiny of figures 1 and 2 of [26], to be true only for the $X_{max}$ data. The muon data, which are compared with predictions of QGSJET98, look to be consistent with a constant and heavy mass from 5 x $10^{16}$ to beyond $10^{18}$ eV. It would be very interesting to establish that the same model gives different predictions for the mass variation with energy for different measured quantities: this might lead to further understanding of the models. However, the systematic errors in measurements of $X_{max}$, as discussed in §3.3 above, may also be a serious problem.

The above discussion demonstrates the difficulties with which one is faced with when trying to compare data. The measurements from different experiments are rarely analysed contemporaneously and the shifts in the inferences from the use of different models can be substantial.

## 4. Possibilities of Identifying Photon and Neutrino Primaries

**4.1. Super-heavy relic particles:** An idea to explain the UHECRs that have been reported beyond 100 EeV is that super-heavy relic particles with masses of ~$10^{12}$ GeV, produced in the early Universe, may decay to produce high energy cosmic rays[27]. While details of the fragmentation of these particles remains a matter of debate, it is generally accepted that the resulting UHECR beam would contain copious fluxes of neutrinos and photons.

**4.1. Limits to the fraction of photon primaries:** It is unlikely that the majority of the events claimed to be near $10^{20}$ eV have photons as parents as some of the showers seem to have normal numbers of muons, the tracers of primaries that are nuclei[a], (see paper in these proceedings by Shinosaki).

___________________

[a] This assumes that the photo-pion production cross-section behaves 'normally'

Furthermore, the cascade profile of the most energetic fluorescence event is inconsistent with that of a photon primary[28]. An alternative method of searching for photons has recently been developed using showers incident at very large zenith angles. Deep-water tanks have a good response to such events out to beyond 80º. At such angles the bulk of the showers detected are created by baryonic primaries but they are distinctive in that the electromagnetic cascade stemming from neutral pions has been almost completely suppressed by the extra thickness of atmosphere penetrated. At 80° the atmospheric thickness is ~ 5.7 atmospheres. At Haverah Park, such large zenith angle showers were observed and the shower disc was found to have a very small time spread. A complication for the study of inclined showers is that the muons, in their long traversal of the atmosphere, are severely bent by the geomagnetic field. A study of this has been made and it has been shown that the rate of triggering of the Haverah Park array at large angles can be predicted[29]. In addition, it was found that the energy of the primaries could be estimated with reasonable precision so that an energy spectrum could be derived. The concept of using the known and mass independent, spectrum deduced by the fluorescence detectors to predict the triggering rate as a function of the mass of the primary has led to a demonstration that the photon flux at $10^{19}$ eV is less than 40% of the baryonic component[30], a conclusion similar to that of the AGASA group, searching for showers which have significantly fewer muons than normal[31].

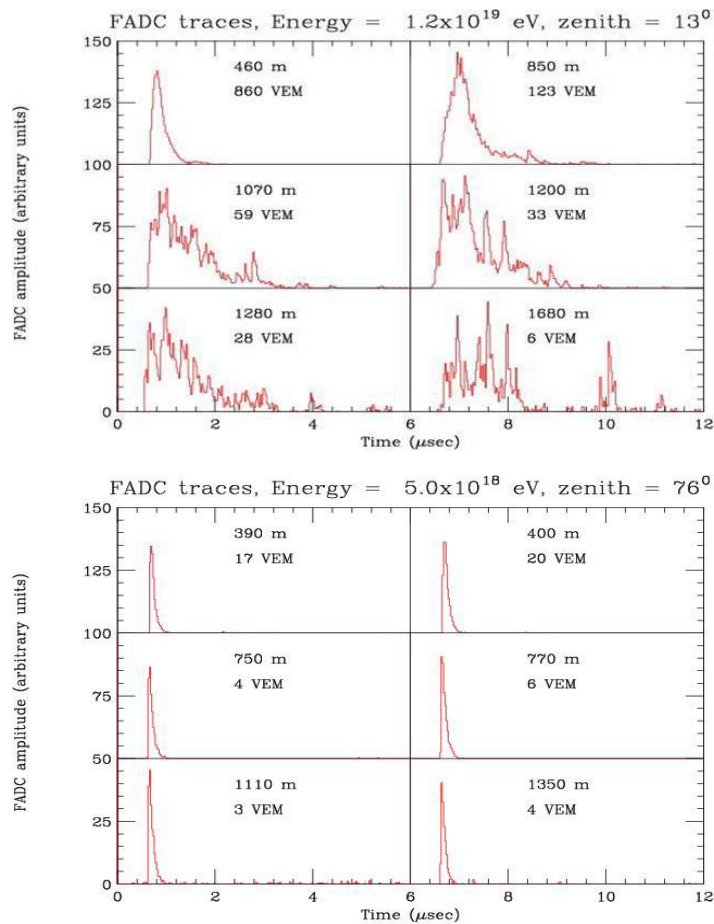

**Figure 4:** The upper plot shows FADC traces from a 13° shower: the radius of curvature was 4 km, and the density ratio 134 over a distance ratio of 3.7. By contrast the lower figure shows a shower of 76° which has a radius of curvature of 27 km and a density ratio of 7.5 for a distance ratio of 3.5. (Picture from Auger Collaboration: made by J W Cronin[38]).

The flux of photons expected at $10^{19}$ eV has recently been reassessed[32] and is expected to be lower than originally predicted. This arises because of a more detailed examination of the fragmentation process. A large flux of photons is now predicted at 100 EeV and can be sought in the Auger data using the method developed at Haverah Park.

**4.2. Neutrino Primaries:** Neutrino primaries may be detectable by studying very inclined showers. This idea was first proposed by Berezinsky and Smirnov[33] and was re-examined in the context of the Auger Observatory by Capelle et al.[34]. A neutrino can interact anywhere in the atmosphere with equal probability. However, if one restricts a search to large zenith angles then it should be possible to identify occasions when the neutrino has interacted in the air mass above the detector. In the case of the 3000 km$^2$ of the Auger detector, the air volume up to 1 km above the detector contains about 3 km$^3$ water equivalent. A neutrino induced shower arriving at a large zenith angle will have distinctive characteristics. The particles will be spread out in time, the signal size will fall off rapidly from the shower centre and the extreme front of the shower will be very curved. The idea is illustrated in figure 4 using data from the Engineering Array of the Auger Observatory[13] where the signals in a near vertical shower (13º) are compared with one at 76º from the zenith where the shower penetrated ~ 4 times as much matter. The broad FADC traces, the curved shower front (4 km as compared with 27 km) and the steep fall off of signal size seen in the near vertical event are the signatures that will be sought in inclined events and, if any are found, used to assess the events as neutrino candidates. It is not clear that there will be sufficient neutrinos to be detectable. The $\nu_\mu$ and $\nu_e$ from the GZK processes probably have too low a flux though $\nu_\tau$, expected in the primary beam because of neutrino oscillations, may be observed[35] as might be neutrinos from some models for AGNs[36] and GRBs[37].

**5. Conclusions:** To make full use of forthcoming information on the energy spectrum and arrival direction distribution at the highest energies, and to interpret what already exists, it is necessary to improve our knowledge of the mass of the cosmic rays above $10^{19}$ eV. Such evidence as there is does not support the widely adopted assumption that all of these cosmic rays are protons: there may be a substantial fraction of iron nuclei present, even at $10^{20}$ eV. Photons do not appear to dominate at $10^{19}$ eV

**6. Acknowledgements:** I would like to thank Antonio Insolia and the organisers of the CRIS meeting for inviting me to Catania and for financial support. Lively discussions with Professor Tere Dova sharpened some of the arguments of §3.6. Work on UHECR at the University of Leeds is supported by PPARC, UK